\documentclass[prl,twocolumn,showpacs,superscriptaddress]{revtex4}
\usepackage[dvips]{graphicx}
\usepackage{color}
\usepackage{amsmath}
\usepackage{time}

\newcommand{\h}{{\mathbf H}}

\newcommand{\expect}[1]{\langle \Phi | #1 | \Phi \rangle}

\newcommand{\ket}[1]{| #1 \rangle}
\newcommand{\bra}[1]{\langle #1 |}

\newcommand{\Ad}[1]{{\mathbf a}^\dag_{#1} }
\newcommand{\A}[1]{{\mathbf a}_{#1} }

\begin{document}

\preprint{LA-UR-05-0557}

\title
   {Density and spin response functions in ultracold fermionic atom gases}
\author{Bogdan~Mihaila}
\affiliation{Theoretical Division,
   Los Alamos National Laboratory,
   Los Alamos, NM 87545}
\author{Sergio~Gaudio}
\affiliation{Department of Physics,
   Boston College,
   Chestnut Hill, MA 02167}
\affiliation{Theoretical Division,
   Los Alamos National Laboratory,
   Los Alamos, NM 87545}
\author{Krastan~B.~Blagoev}
\affiliation{Theoretical Division,
   Los Alamos National Laboratory,
   Los Alamos, NM 87545}
\author{Alexander~V.~Balatsky}
\affiliation{Theoretical Division,
   Los Alamos National Laboratory,
   Los Alamos, NM 87545}
\author{Peter~B.~Littlewood}
\affiliation{Cavendish Laboratory,
   Madingley Road,
   Cambridge CB3 0HE,
   United Kingdom}
\author{Darryl~L.~Smith}
\affiliation{Theoretical Division,
   Los Alamos National Laboratory,
   Los Alamos, NM 87545}


\begin{abstract}
We propose a new method of detecting the onset of superfluidity in a
two-component ultracold fermionic gas of atoms governed by an
attractive short-range interaction. By studying the two-body
correlation functions we find that a measurement of the momentum
distribution of the density and spin response functions allows one
to access separately the normal and anomalous densities. The change
in sign at low momentum transfer of the density response function
signals the transition between a BEC and a BCS regimes,
characterized by small and large pairs, respectively. This change in
sign of the density response function represents an unambiguous
signature of the BEC to BCS crossover. Also, we predict spin
rotational symmetry-breaking in this system.
\end{abstract}


\maketitle

The BEC-to-BCS crossover has drawn renewed interest in recent years
due to experimental progress in atom trap systems~\cite{traps}.
Several groups~\cite{mol-fermi_BEC} achieved a Bose-Einstein
condensate (BEC) in which the fermions form non-overlapping
(Shafroth) pairs~\cite{small-pairs} in a two-component Fermi gas.
Despite considerable effort, much work is still needed to understand
the other limit in which the pairs are large, and the pairing occurs
in momentum space, similarly to the Bardeen-Cooper-Schrieffer (BCS)
state of superconductivity in normal metals. Questions still remain
regarding whether or not the system is superfluid when passing
through the crossover with temperatures of the order of twenty
percent of the Fermi temperature. The appropriate characterization
of the system is still open to debate. Recently, several groups have
claimed~\cite{jin,ketterle} to have reached the superfluid state on
the negative scattering length side of the Feshbach resonance.
Despite evidence of fermionic pairing, these claims are still
subject to intense discussions as no definitive proof of
superfluidity~\cite{ketterle} is available yet.

Viverit \emph{et al.}~\cite{momentum} have suggested recently that
the shape of the atomic momentum distribution at low temperatures is
very sensitive to the sign and size of the scattering length. Altman
\emph{et al.} have proposed to utilize density-density correlations
in the image of an expanding gas cloud to probe complex many-body
states of trapped ultracold atoms. Also, Bruun and Baym~\cite{baym}
showed that scattered light from a fermionic gas would exhibit a
large maximum below the superfluid critical temperature and
therefore it can be used to detect the superfluid transition.
However it is not clear how this behavior changes close to the
crossover, where the actual experiments are performed.

In this paper we propose a new diagnostic method for fermionic
atomic gas condensates. By studying the zero-temperature evolution
of the density and spin response functions, as a function of the
scattering length of the interaction, we show that the density
response function changes sign across the BEC to BCS crossover and
that this change can be used to experimentally distinguish the BEC
from the BCS state of the system.


Our model system consists of fermionic atoms in two hyperfine states
interacting via a finite-range attractive interaction. Based on a
variational approach, we also derive a sum-rule satisfied by the
spin-spin correlation function at $q=0$. We find that the density
and spin response functions are given by the sum and the difference
of independent contributions arising from the normal and anomalous
density, respectively. By measuring the momentum distribution of the
two response functions it is possible to infer separately the normal
and anomalous densities. Finally, we predict symmetry breaking of
the spin-rotation invariance. This prediction is characteristic to
our model and, therefore, provides an experimental check for its
applicability to real systems.

\begin{figure}[t!]
   \includegraphics[width=2.75in]{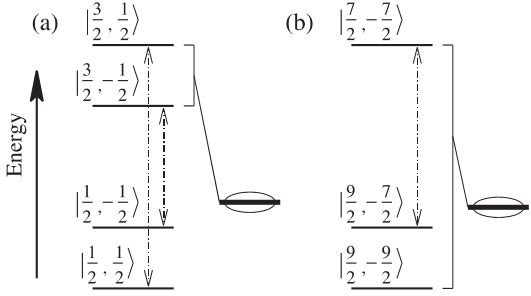}
   \caption{
   \label{fig:Fesh}
Atomic states involved in the Feschbach resonance at small magnetic
field, $B$, for (a) ${}^{6}$Li and (b) ${}^{40}$K. The hyperfine
couplings between states allowed by the selection rules are
represented by dotted lines. In ${}^6$Li, the bound state is formed
from the $F=3/2$ states while, in ${}^{40}K$, it is formed from
$\ket{\frac{7}{2}\, -\frac{7}{2}}$ and the lowest eigenstate
$\ket{\frac{9}{2}\, -\frac{9}{2}}$.}
\end{figure}

We begin by considering the two-body Hamiltonian
\begin{align}
   \h =
   \epsilon_{k\!, i} \, \Ad{k i} \,\A{k\! i}
   +
   \frac{1}{2} \,
   V_{q; i m, n j} \,
   \Ad{k m} \,\Ad{p i} \,
   \A{p - q j} \,\A{k + q \, n}
\label{eq:ham_0}
   \>,
\end{align}
where $\{\Ad{k i}, \A{k i}\}$ are the \emph{particle} creation and
annihilation operators corresponding to single-particle states of
linear momentum $\mathbf{k}$ and fermion type $i$. The atomic
levels, which we associate with the fermion types, are eigenstates
of the Hamiltonian of an ion (with integer nuclear spin $I$)
interacting with an electron (spin $s=\frac{1}{2}$)
\begin{equation}
   \label{Vatom}
   \hat{H}_{\rm atom} = k.e. \ + \ A\, \vec{s}\cdot \vec{I}+\vec{B}\cdot
   \bigl ( 2\,\mu_{\rm e}\, \vec{s}-\mu_{\rm n}\, \vec{I} \bigr )
   \>,
\end{equation}
where $A$ denotes the strength of the hyperfine interaction and
$\vec{B}$ is the magnetic field, while $\mu_{\rm e}$ and $\mu_{\rm
n}$ denote the electron and nuclear magnetic moments, respectively.
We introduce the total angular momentum, $\vec F= \vec I + \vec s$.
The total angular momentum projection, $M_F$, is the only good
quantum number at finite $B$. The atomic spectra considered in our
model are depicted in Fig.~\ref{fig:Fesh}.


In the Hartree-Fock-Bogoliubov (HFB) formalism, one first introduces
the \emph{quasi-particle} creation and annihilation operators, $\{
\beta_{k i}, \beta_{\overline k i} \}$, in terms of the
\emph{particle} creation and annihilation operators, via the
Bogoliubov-Valatin transformation~\cite{BV}. Then, the ground state,
$\ket{\Phi}$, of the Hamiltonian~\eqref{eq:ham_0}, is obtained via a
variational ansatz, such that
$
   \beta_{k i} \ket{\Phi} = 0
$, 
for all $\{ki\}$ labels. We will refer to this state as the HFB wave
function.


While the above approach is general for any arbitrary multi-level
fermionic Hamiltonian~\cite{ref:3level}, for the purpose of the
present discussion we will confine ourselves to a two level
(one-channel) model, such as discussed
in~\cite{ref:CN82,ref:2level}. In this model the HFB wave function
has the BCS form
\begin{equation}
   | \Phi \rangle \ = \
   \prod_{\mathbf{k}} \ \left ( u_{\mathbf{k}} \ + \ v_{\mathbf{k}} \
   a^{\dag}_{\mathbf{k} \uparrow} a^{\dag}_{-\mathbf{k} \downarrow}
   \right ) \ | 0 \rangle
\label{eq:BCS_gs}
   \>,
\end{equation}
subject to the normalization condition,
\begin{equation}
   | u_{\mathbf{k}}|^2 + |v_{\mathbf{k}}|^2 = 1
   \>.
\label{eq:uv_norm}
\end{equation}


\begin{table}[b!]
   \caption{Parameters entering the modified Pauli matrices.}
   \begin{tabular}{lccccccc}
      \\
        Atom &
        a & b & c & d & $\ (b^2-a^2)$ & $\ (d^2-c^2)$ & bc
      \\
      \colrule
      [B = 0] \\
      $^6$Li &
      $\frac{1}{\sqrt 3}$ & $\sqrt{ \frac{2}{3} }$ & $\sqrt{ \frac{2}{3} }$ & $\frac{1}{\sqrt 3}$ &
      $\frac{1}{3}$ & $-\frac{1}{3}$ & $\frac{2}{3}$
      \\
      $^{40}$K &
      $\frac{1}{3}$ & $- \frac{2\sqrt{2}}{3}$ & 1 & 0 &
      $\frac{7}{9}$ & $- 1$ & $- \frac{2\sqrt{2}}{3}$
      \\
      \colrule
      [$B \rightarrow \infty$] \\
      $^6$Li &
      0 & 1 & 0 & 1 &
      1 & 1 & 0
      \\
      $^{40}$K &
      1 & 0 & 1 & 0 &
      -1 & -1 & 0
      \\
      \colrule
   \end{tabular}
\label{tab:sigmas}
\end{table}


For zero magnetic field, $B$, the states $\ket{\uparrow}$ and
$\ket{\downarrow}$ refer to the two fermion types $\ket{F\, M_F}
\equiv \ket{\frac{3}{2},\frac{1}{2}}$ and
$\ket{\frac{3}{2},-\frac{1}{2}}$ for $^{6}$Li, and
$\ket{\frac{7}{2},-\frac{7}{2}}$ and
$\ket{\frac{9}{2},-\frac{9}{2}}$ for $^{40}$K. For finite values
of~$B$, we have
\begin{align}
   ``\ket{\uparrow}''
   & =
   a\, \ket{11} \ket{\downarrow} + b\, \ket{10} \ket{\uparrow}
   \quad \stackrel{\tiny B\rightarrow 0}{\longrightarrow} \quad
   \textstyle{ \ket{\frac{3}{2} \frac{1}{2}} }
\label{eq:li6_spins}
   \>,
   \\ \notag
   ``\ket{\downarrow}"
   & =
   c\, \ket{10} \ket{\downarrow} + d\, \ket{1-1} \ket{\uparrow}
   \quad \stackrel{\tiny B\rightarrow 0}{\longrightarrow} \quad
   \textstyle{ \ket{\frac{3}{2} -\frac{1}{2}} }
   \>,
\end{align}
for $^6$Li, while for $^{40}$K we have
\begin{align}
   ''\ket{\uparrow}''
   & =
    a\, \ket{4~-3} \ket{\downarrow} + b\, \ket{4~-4} \ket{\uparrow}
   \quad \stackrel{\tiny B\rightarrow 0}{\longrightarrow} \quad
   \textstyle{ \ket{\frac{7}{2} - \frac{7}{2}} }
   \>,
   \notag \\
   ``\ket{\downarrow}"
   & =
   \ket{4~-4} \ket{\downarrow}
   \quad \stackrel{\tiny B\rightarrow 0}{\longrightarrow} \quad
   \textstyle{ \ket{\frac{9}{2} -\frac{9}{2}} }
\label{eq:k40_spins}
   \>.
\end{align}
The parameters ($a,b,c,d$) are related to the hyperfine mixing
angles, such that
\begin{align}
   a = \sin \phi_1\>, \ b = \cos\phi_1\>, \
   c = \sin \phi_2\>, \ d = \cos\phi_2\>,
\end{align}
for $^6$Li, and
\begin{align}
   a = \cos \phi_1\>, \ b = \sin \phi_1\>, \
   c = \cos \phi_2\>, \ d = \sin \phi_2\>,
\end{align}
for $^{40}$K. When the magnetic field is zero, the above parameters
are given by the appropriate Clebsch-Gordan coefficients, while for
large fields we retrieve the unmixed phase, as illustrated in
Table~\ref{tab:sigmas}.


The ground-state properties are described by the \emph{normal} and
\emph{anomalous} densities defined as
\begin{align}
   \rho_{\mathbf{k}}
   & \ = \
      \expect{ \Ad{\mathbf{k} \uparrow} \A{\mathbf{k} \uparrow} }
   \ = \ |v_{\mathbf{k}}|^2
   \>,
\label{eq:rho_def}
   \\
   \kappa_{\mathbf{k}}
   & \ = \
      \expect{ \A{-\mathbf{k} \downarrow} \A{\mathbf{k} \uparrow} }
   \ = \ v_{\mathbf{k}}^* u_{\mathbf{k}}
   \>,
\label{eq:kappa_def}
\end{align}
while the mean total particle-density of the system, $\rho_0$, is
given by
\begin{equation}
   \rho_0 \ = \ \langle \Phi | \hat{N} | \Phi \rangle
   \ = \ 2 \ \int \ \frac{\mathrm{d^3}k}{(2\pi)^3}  \ \rho_{\mathbf{k}}
   \>.
\label{eq:no}
\end{equation}

\begin{figure}[t!]
   \includegraphics[width=0.38\textwidth]{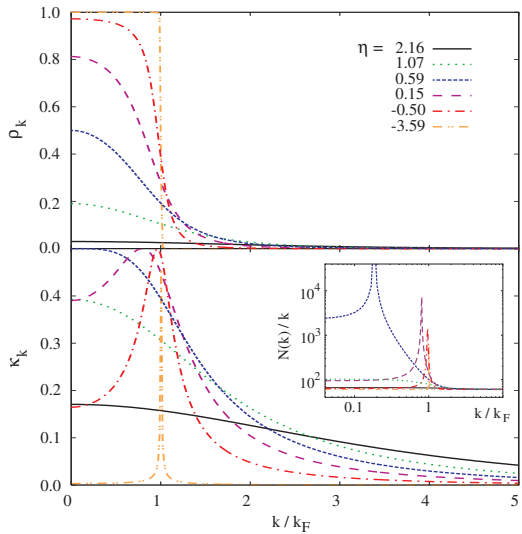}
   \caption{\label{fig:densities}
   (Color online)
Momentum distribution of the normal and anomalous densities, as a
function of the scattering length, at fixed system density, $\rho_0$
($k_F \langle r \rangle \approx0.37$)~. Inset shows the density of
states. We use the notation $\eta = (a_{0}k_{F})^{-1}$.}
\end{figure}

The ground-state ansatz~\eqref{eq:BCS_gs} provides a smooth
interpolation between the BCS and the BEC regimes. We have
recently~\cite{ref:2level} used this model to study the properties
of the BEC-BCS crossover, for a short- (but finite) range and
attractive interaction. In the dilute limit the model is equivalent
to the zero-range (contact) interaction Hamiltonian, initially
discussed by Leggett~\cite{ref:leg80}. The study presented in
Ref.~\cite{ref:2level} was carried out by modifying the scattering
length of the interaction, while keeping the density and range of
the interaction fixed. Figure~\ref{fig:densities} illustrates the
momentum distributions of the normal and anomalous densities. The
mean-field solution predicts the crossover occurs when the minimum
in the quasi-particle energy spectrum shifts from a finite (BCS) to
zero-momentum value (BEC). As such, the presence/absence of the
singularity in the momentum distribution of the density of states
represents a unambiguous signature of the crossover (see inset in
Fig.1).


Here, we focus on the changes in the two-point correlation functions
as we evolve the system from the BEC to the BCS regimes. We begin by
defining the
response functions
\begin{align}
   S_{\rho}(q) =
   \frac{1}{\rho_0}
   \int [\mathrm{d}^3\xi] &
   e^{\mathrm{i} \mathbf{q} \cdot \xi}
   \expect{\rho(\mathbf{r}) \rho(\mathbf{r}')}|_{\xi = \mathbf{r} - \mathbf{r}'}
   \>,
\label{eq:S_el}
   \\
   S^{\mu \nu}_{\sigma}(q) =
   \frac{1}{\rho_0}
   \int [\mathrm{d}^3\xi] &
   e^{\mathrm{i} \mathbf{q} \cdot \xi}
   \expect{S_\mu(\mathbf{r}) S_\nu(\mathbf{r}')}|_{\xi = \mathbf{r} - \mathbf{r}'}
\label{eq:S_mag}
   \>,
\end{align}
where $\rho(\mathbf{r})$ and $S_\mu(\mathbf{r})$ denote the
particle- and spin-density operators
\begin{align}
   \rho(\mathbf{r}) &
   =
\label{eq:particle_den}
   \sum_{ij}
   \Ad{i}
   \langle i | \delta(\mathbf{r} - \mathbf{r}_1)
   | j \rangle \
   \A{j}
   \>,
   \\
   S_\mu(\mathbf{r}) &
   =
\label{eq:spin_den}
   \frac{1}{2}
   \sum_{ij}
   \Ad{i}
   \langle i | \sigma_{\mu} \, \delta(\mathbf{r} - \mathbf{r}_1)
   | j \rangle
   \A{j}
   \>.
\end{align}
Here, $\sigma_{\mu}$ denote the Pauli matrices. Since the
single-particle states are plane waves, i.e.
$
   \ket{i}
   \ = \
   e^{\mathrm{i} \mathbf{k}_i \cdot \mathbf{r}}
   \chi_i
   \>,
$
where $\chi$ denotes the ($\ket{\uparrow}$,
$\ket{\downarrow}$) spinors, then we can calculate the
particle-density matrix element as
\begin{align}
   \langle i | \delta(\mathbf{r} - \mathbf{r}_1)
   | j \rangle
   \ = \
   \delta_{ij} \
   e^{\mathrm{i} ( \mathbf{k}_j - \mathbf{k}_i ) \cdot \mathbf{r}}
   \>,
\end{align}
and the spin-density matrix element as
\begin{align}
   \langle i | \sigma_{\mu} \, \delta(\mathbf{r} - \mathbf{r}_1)
   | j \rangle
   \ = \
   \bra{\chi_i} \sigma_\mu \ket{\chi_j} \
   e^{\mathrm{i} ( \mathbf{k}_j - \mathbf{k}_i ) \cdot \mathbf{r}}
\label{eq:spin_me}
   \>.
\end{align}


\begin{figure}[t!]
   \includegraphics[width=0.385\textwidth]{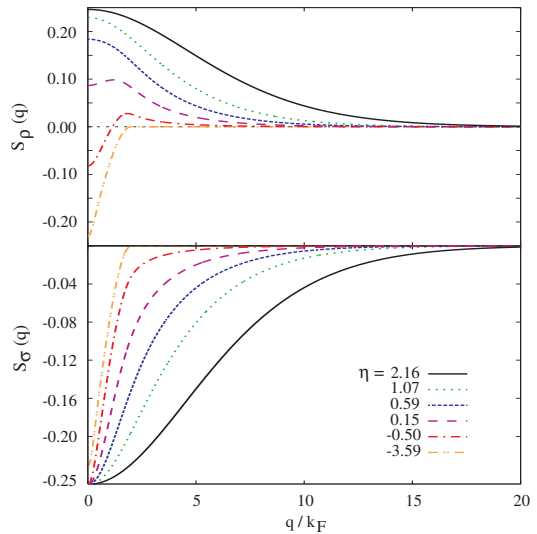}
   \caption{\label{fig:inst_resp}
   (Color online)
The density and spin instantaneous response function for an electron
gas, as a function of the scattering length (or $\eta =
(a_{0}k_{F})^{-1}$), at fixed density,~$\rho_0$.}
\end{figure}


The density response function in~\eqref{eq:S_el} is a scalar, while
the spin density response function in~\eqref{eq:S_mag} is a tensor.
For the one-channel model the spin-response tensor is diagonal, and,
provided that the Hamiltonian and the total spin operator commute,
the diagonal components are equal. Then, the density and spin
response functions correspond to the sum and difference of
\emph{separate} normal and anomalous density contributions,
respectively. The nonlocal response functions can be written as
\begin{align}
   S_\rho(q)
   \ \doteq \ &
\label{eq:sel_static}
   \mathcal{I}_\kappa(q) \ - \ \mathcal{I}_{\rho}(q)
   \>,
   \\
   S_\sigma(q)
   \ \doteq \ &
\label{eq:smag_static}
   - \
   \frac{1}{4} \
   \Bigl [
   \mathcal{I}_\kappa(q) \ + \ \mathcal{I}_{\rho}(q)
   \Bigr ]
   \>.
\end{align}
Equations~\eqref{eq:sel_static} and~\eqref{eq:smag_static} represent
a general result for the HFB mean-field approximation of the
ground-state of the multi-level Hamiltonian~\eqref{eq:ham_0}. For
the one-channel model, and disregarding for now the hyperfine nature
of the atomic levels involved (similarly to the case of an electron
gas), the normal and anomalous density contributions to the response
functions are:
\begin{align}
   \mathcal{I}_{\rho}(q)
   = &
   \frac{2}{\pi^3\rho_0}
   \int_0^\infty \!\!\!\! \mathrm{d}\xi \, \xi^2 j_{0}(\xi q)
   \left [
   \int_0^\infty \!\!\!\! \mathrm{d}k \, k^2 \rho_{k} j_{0}(\xi k)
   \right ]^2
   \>,
   \\
   \mathcal{I}_{\kappa}(q) = &
   \frac{2}{\pi^3\rho_0}
   \int_0^\infty \!\!\!\! \mathrm{d}\xi \, \xi^2 j_{0}(\xi q)
   \left [
   \int_0^\infty \!\!\!\! \mathrm{d}k \, k^2 \kappa_{k} j_{0}(\xi k)
   \right ]^2
   \>.
\end{align}


The response functions $S_\rho(q)$ and $S_\sigma(q)$ are shown in
Fig.~\ref{fig:inst_resp}. The density response function,
$S_\rho(q)$, changes sign at low momentum, as the bound state
disappears, 
while the spin response function, $S_\sigma(q)$, changes smoothly
from the BEC to the BCS regime. In the BCS limit, the condensate
wave function, $\kappa_k$, appears in a very narrow region around
the Fermi momentum (see Fig.~\ref{fig:densities}). Since the width
of this distribution tends to zero in the limit $a_0 \rightarrow -
0$, so does the anomalous density contribution,
$\mathcal{I}_{\kappa}(q)$. In turn, the density and spin response
functions will be equal in this limit.

\begin{figure}[t!]
   \includegraphics[width=0.385\textwidth]{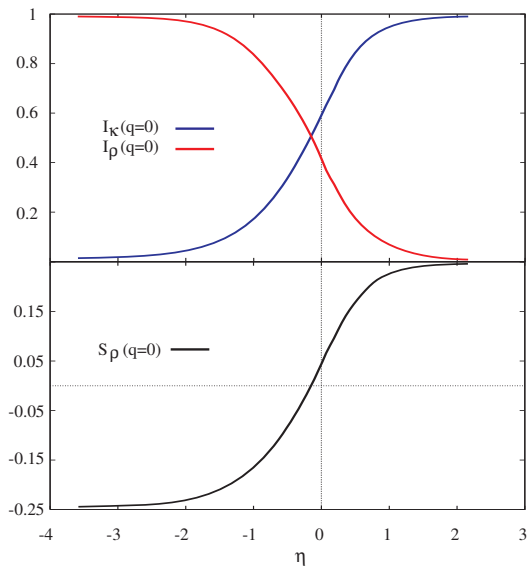}
   \caption{\label{fig:zero_q}
   (Color online)
Zero-momentum transfer normal and anomalous density contributions,
$\mathcal{I}_\kappa(0)$ and $\mathcal{I}_\rho(0)$, together with the
density response function $S_\rho(q)$ for $q=0$.}
\end{figure}

The mean-field approach shows that a measurement of the density
response function is a signature of the BEC-BCS crossover. In the
dilute limit, the crossover coincides with the singularity in the
scattering length, which in turn corresponds to the change in sign
of the density response function. Figure~\ref{fig:zero_q}~(top)
shows the dependence of $\mathcal{I}_\kappa(q)$ and
$\mathcal{I}_\rho(q)$ for $q=0$, as a function of the scattering
length $a_0$ at fixed density $\rho_0$ ($k_F \langle r \rangle
\approx0.37$). The corresponding density response function
$S_\rho(0)$ is depicted in Fig.~\ref{fig:zero_q}~(bottom). The
density response function changes sign 
between the BCS and the BEC limits.

The one-channel model predicts that the spin response function at
zero momentum transfer, $S_\sigma(q=0)$, is in fact a sum rule, i.e.
\begin{align}
   S_\sigma(0) = &
   -
   \frac{1}{4\pi^2\rho_0}
   \int_0^\infty \mathrm{d}k \ k^2 \
   ( \kappa_{k}^2 + \rho_{k}^2 )
   =
   - \,
   \frac{1}{4}
   \>,
\end{align}
or $\mathcal{I}_\rho(0) + \mathcal{I}_\kappa(0) = 1$ is independent
of $\eta = (a_{0}k_{F})^{-1}$, as shown in Fig.~\ref{fig:zero_q}
(top). The above is obtained by using the
definitions~\eqref{eq:rho_def} and~\eqref{eq:kappa_def}, together
with the normalization condition, Eq.~\eqref{eq:uv_norm}.


We now consider the modification of the electron gas results, due to
the hyperfine nature of the interacting atomic levels. We find the
spin response function is proportional to the ``weighted''
difference of the normal and anomalous density contributions derived
earlier, $\mathcal{I}_\kappa(q)$ and $\mathcal{I}_\rho(q)$. The
weighting factors depend on the atom specie in the system, and the
atomic levels involved in the interaction. Since the spin response
function is only sensitive to the electron spin operator, then in
order to find the modification of the spin response function, we
need to calculate the matrix elements $\bra{\chi_i} \sigma_\mu
\ket{\chi_j}$ in Eq.~\eqref{eq:spin_me}. The parameters entering the
modified Pauli matrices are
\begin{align}
   \sigma_0
   \ \rightarrow \
   \left (
   \begin{array}{cc}
   b^2 - a^2 & 0 \\
   0 & d^2 - c^2
   \end{array}
   \right )
   \>, \quad
   \sigma_{\pm} \ \rightarrow \ bc \ \sigma_{\pm}
   \>,
\end{align}
and limiting values as a function of the magnetic field, $B$, are
listed in Table~\ref{tab:sigmas}. We note that in the large field
limit, the spin response in a $^6$Li fermionic atom gas has opposite
sign as compared to the case of a $^{40}$K fermionic atom gas.
Irrespective of the actual B value, we obtain that for a fermionic
atom gas, the rotational symmetry of the spin response function is
broken. This symmetry-breaking effect is a prediction of our model,
and arises as a consequence of the fact that our effective
interaction involves a restricted set of atomic levels.


In conclusion, in this paper, we have shown that much information
about the crossover regime can be gained by experimentally studying
the density and spin response functions. Within the framework of the
mean-field results, we show that the normal and anomalous densities
can be accessed from the momentum distribution of the response
functions. The spin response function changes smoothly across the
crossover, while the density response function changes sign, and
thus represents a signature of the crossover. The spin response at
zero-momentum transfer satisfies a sum rule, and is sensitive to the
interaction. When taking into account the hyperfine structure of the
interacting levels, our model predicts that the rotational
invariance of the spin response, normally associated with an
electron gas subject to a spin-independent interaction, is broken.




%
%
%
%
%

\end{document}